\newif\ifOSA
\OSAtrue

\ifOSA
\documentclass{osa-article}
\journal{oe}
\articletype{Research Article}

\fi

\usepackage{amsmath,amssymb}

\usepackage{graphicx}

\begin{document}

\ifOSA

\title{Shaping Coherent X-rays with Binary Optics}

\author{Stefano Marchesini\authormark{1} }
\author{Anne Sakdinawat\authormark{2} }

\address{\authormark{1}
Computational Research Division, Lawrence Berkeley National Laboratory, 1 Cyclotron Road, Berkeley, CA, 94720, USA
\\
\authormark{2}
SLAC National Accelerator Laboratory, 2575 Sand Hill Road, Menlo Park, CA, 94025, USA
}

\email{\authormark{*} smarchesini@lbl.gov, annesak@stanford.edu}

\begin{abstract}
    Diffractive lenses fabricated by lithographic methods are one of the most popular image forming optics in the x-ray regime.  Most commonly, binary diffractive optics, such as Fresnel zone plates are used due to their ability to focus at high resolution and to manipulate the x-ray wavefront.  We report here a binary zone plate design strategy to form arbitrary illuminations for coherent multiplexing, structured illumination, and wavefront shaping experiments.  Given a desired illumination, we adjust the duty cycle, harmonic order, and zone placement to vary both the amplitude and phase of the wavefront at the lens.  This enables the binary lithographic pattern to generate arbitrary structured illumination optimized for a variety of applications such as holography, interferometry, ptychography, imaging, and others.
\end{abstract}

\fi


\section{Introduction}
High brightness synchrotron light sources and x-ray free electron lasers \cite{kim2017x, schmuser2014fel, eriksson2014dlsr} are currently being built around the world due to their potential for new x-ray science \cite{attwood2017x} opportunities, ranging from materials to biological sciences.  One of the primary advantages of these sources is the large amount of spatially coherent x-rays that can be generated.  The use of these coherent x-rays \cite{Nugent2010coherentmethods}, combined with wavefront-shaping diffractive optics, can open up new possibilities for x-ray measurements in imaging, spectroscopy, and scattering.  However, one of the primary challenges associated with arbitrary wavefront-shaping diffractive optics for the x-ray regime is the difficulty of fabricating 3D structures that account for both the amplitude and phase modifications needed for accurate wavefront shaping.  Current methods for designing these optics rely on either simple phase thresholding, which works well for basic analytical cases \cite{Sakdinawat2007spiral} but becomes inaccurate for arbitrary shaping, or iterative techniques \cite{lee1979binary, fienup1980iterative}, which have limitations in accuracy as well. Alternatively, binary photon sieves \cite{xie2012toward,xie2010feasibility} have been proposed for nano-focusing but they are also unable to produce arbitrary shaping. Overcoming these challenges requires a new diffractive optics design strategy that provides accurate shaping while falling within the practical constraints of current nanofabrication capabilities.  

To produce high-fidelity structured illumination while maintaining a binary design, three new diffractive optics design concepts have been developed and are presented here.  Given a desired illumination, various parameters such as duty cycle, harmonic order, and zone placement are adjusted to vary both the amplitude and phase of the wavefront at the lens.  These binary diffractive optics can then be used to generate arbitrary structured illumination optimized for a variety of applications such as holography \cite{marchesini2008massively}, imaging \cite{DiFabrizio2003doedic,di2003shaping,Chang2006diczp,vonHofsten2008dictheory}, interferometry, ptychography\cite{stockmar2013near,morrison2018x,chen2018coded,marchesini2016alternating,maiden2013soft}, and other coherent multiplexing experiments.

\section{Binary diffractive optics with amplitude modulation}

\begin{figure}[h]
\centering \includegraphics[width=0.4\columnwidth]{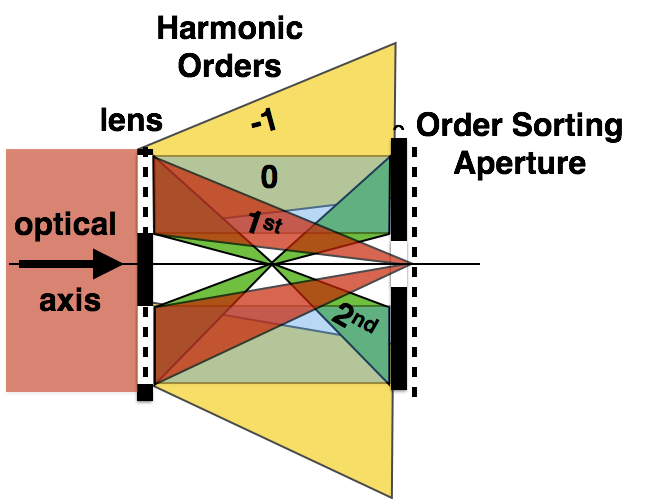}    \includegraphics[width=0.4\columnwidth]{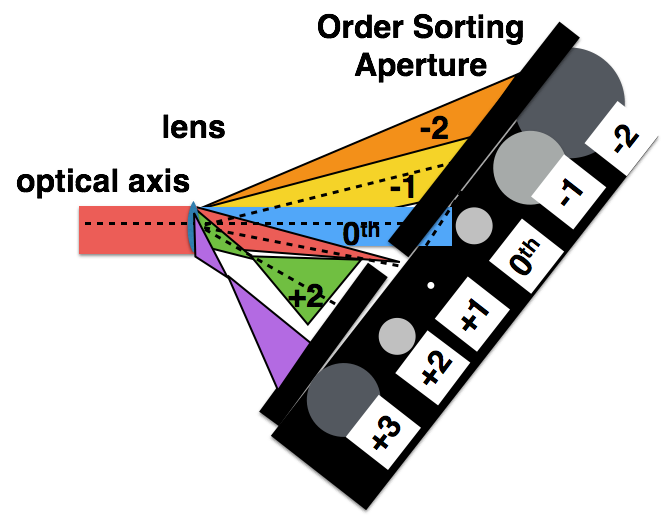}
    \caption{On axis and off axis Zone plates with order sorting apertures}
    \label{fig:my_label}
\end{figure}

The standard Fresnel zone plate is obtained by a binary threshold of the phase of a propagated point source.  This can be done either in an on-axis \cite{Gabor1948holography} case or an off-axis \cite{leith1962offaxis} case, as shown in figure \ref{fig:my_label}.  In combination with an order sorting aperture, the desired focal spot can be selected and utilized in an experiment.  On-axis zone plates have the advantage of maintaining higher resolution focusing for a given outermost zone width while off-axis zone plates have the advantage of cleaner separation among the diffractive orders.

Given the wavelength $\lambda$, and a focal distance $f$, the wavefront to achieve focusing is proportional to $e^{i \alpha r^2}$,  with $\alpha=\frac{\pi}{f \lambda}$ and $r$ the distance from the optical axis.  Setting $\phi_0=\alpha r^2$, the phase of the wavefront, the zone plate binary pattern $\mathrm{ZP}_{0}$  is expressed as:

\begin{equation}
{\mathrm{ZP}_0}(\phi_0)=
\left \{ \begin{array}{l l}
    1, \text{if  $\quad \mathrm{mod}\left (\phi_0,2 \pi \right )< \pi$} \\
    0, \text{otherwise,} 
\end{array} \right .
\end{equation}
which can be viewed as a binarized version of the the point hologram $\Re e^{i \alpha r^2}$.

In order to generate an arbitrary desired illumination $w$, we need to modify the zone plate pattern $\mathrm{ZP}_0$. At the lens plane, a distance $f$ from the illumination plane,  the wavefront desired should resembles $W=e^{i \phi_0} {\cal{F}}w $, where ${\cal{F}}w$ is the Fourier transform of $w$. A commonly used method to generate a diffractive optic to achieve this is to simply apply the same threshold to the phase of the desired wavefront, after adding the carrier phase frequency $\phi_{w}=\mathrm{phase}(W)+\phi_0$. However, this method is prone to quantization errors \cite{lee1974binary} that can severely distort the illumination, and iterative schemes have been proposed since 1980s  \cite{lee1979binary,fienup1980iterative,dixit1994kinoform} to mitigate such artifacts \cite{cheung2011enhanced,TSANG201526}. Unfortunately, the optimization problem is  highly non-linear and non-convex, and such iterative methods may not find an optimal solution that can be practically implemented using current fabrication capabilities.

We propose three methods here for zone plate design that can be used for more accurate wavefront shaping.  The first method locally modifies the amplitude contribution of the wavefront desired at the lens plane W through adjustment of the duty cycle.   The zone plate can be viewed as a grating with variable carrier frequency grating spatial period $p=\alpha r$, $\alpha=\frac{\pi}{f \lambda}$, where the phase satisfies $\phi_0=(\alpha r) r  = p r$. The Fourier coefficient $\varepsilon_h$  of binary pulse train of period $p$  and  $0<L<p$ the non-zero pulse ``linewidth'' is given as:

\begin{equation}
\varepsilon_h(L,p)=\begin{cases}
L/p\text{, if $h=0$}  \\
     \frac{1}{h \pi} \sin \left (
    \frac{\pi h L}{ p} \right ) \text{, if $|h|>0$ }
\end{cases}
\end{equation}
order $h$, where $h$ is an integer.  To  modify the amplitude, we adjust the local efficiency of the zone plate by varying the duty cycle through increasing or decreasing the linewidth. We wish the lens efficiency to match the desired amplitude of the wavefront, 
${\pi} \varepsilon=|W|$, where $|W|$ is normalized so that $|W|\in [0,1]$. 
Define  $\phi_W=\mathrm{asin}(|W|)$ and a binary pattern as:

\begin{equation}
{\mathrm{ZP_{1dc}}}(\phi_w,|W|)=\begin{cases}
    1, \text{if $\mathrm{mod}(\phi_w+\phi_W,2\pi)< 2 \phi_W$,}\\
    0, \text{otherwise}
    \end{cases}
\end{equation}


If the carrier phase factor $\phi_0$ changes rapidly, compared to $\phi_W$, yields a duty cycle $L/p= \tfrac{1}{\pi} \mathrm{asin}(|W|)$, and line-placement shifted by $\mathrm{phase}(W)$.


The practical limitation of this first method is that the line widths in some regions can become too small for fabrication.  In the second method, the fabrication limitations are overcome by replacing those areas in which the linewidths get too small, with a higher harmonic order zone plate region, which have broader linewidths, and fills every other period as:

\begin{equation}
{\mathrm{ZP}_{1h}}(\phi_w,|W|)=\begin{cases} 
    1,  \text{if $\mathrm{mod}(\phi_w+\phi_W,4\pi)< 2 \pi+2 \phi_{W}$}\\
    0, \text{otherwise} 
\end{cases}
\end{equation}

  The hybrid ZP can be generated by combining the two patterns as:

\begin{equation}
{\mathrm{ZP}_{\mathrm{2hyb}}}(\phi_w,|W|)= \begin{cases}
    \text{${\mathrm{ZP}_{1dc}}(\phi_w,|W|)$, if $|W|>1/2$ }\\
    \text{${\mathrm{ZP}_{1h}}(\phi_w,2|W|)$,  otherwise}
\end{cases}
\end{equation}
The combination of the two patterns can be improved by applying a phase unwrapping algorithm to $\phi_w$.

A third method is to use a grating orthogonal to the zone plate with 0 order efficiency proportional to the desired amplitude. The grating has locally variable linewidths that are used to adjust the amplitude contribution in the desired areas.  Such grating-adjusted zone plates can be generated by combining a phase-only zone plate $W_{\mathrm{ZP}_0}$ with a transmission grating $T(|W|)$:

\begin{equation}
{\mathrm{ZP}_{\mathrm{3grat}}}(\phi_w,|W|)=    {\mathrm{ZP}_0}(\phi_w) \circ T(|W|) 
\end{equation}

where the transmission grating has variable linewidth:

\begin{equation}
T(|W|)=\begin{cases} 
    \text{1, if $\mathrm{mod}(x/l-|W|/2,2)<|W|$ }\\
    \text{$0$,  otherwise}
\end{cases}
\end{equation}

where $l$ is the outermost zonewidth.

\section{Simulations}

 
\begin{table}[h]
    \centering
    \begin{tabular}{l|l|l|l|l }
      lens  & nmse  & nmse-o & relative efficiency &$\tfrac{1}{c}$  \\
  \hline
  \hline
  URA\\
  $\|W\|^2=0.0062$\\
  \hline
      ZP$_0$                & 0.550  & 0.550 & 1       & 16    \\
      ZP$_{\mathrm{1dc}}$   & 0.0016    & 0.503 & 0.0072  & 1   \\
      ZP$_{\mathrm{2hyb}}$  & 0.146     & 0.165 & 0.0083  & 1.17  \\
      ZP$_{\mathrm{3grat}}$ & 0.0063    & 0.3092& 0.0072  & 1.0   \\
      \hline
      Holographic\\
      $\|W\|^2=0.0056$\\
      \hline
      ZP$_0$                & 0.648    & 0.649 &  1       &  22    \\
      ZP$_{\mathrm{1dc}}$   & 0.0016   & 0.363 &  0.0056  &  1   \\
      ZP$_{\mathrm{2hyb}}$  & 0.1965   & 0.197 &  0.0074  &  1.3  \\
      ZP$_{\mathrm{3grat}}$ & 0.0016   & 0.265 &  0.0058  &  1.03   \\
       \hline
      Bandwidth Limited\\ Random Ptychography\\
      $\|W\|^2=0.39202$\\
      \hline
      ZP$_0$                & 0.0760  & 0.076   &  1       &  2.5808    \\
      ZP$_{\mathrm{1dc}}$   & 0.00031 &  0.0019 &  0.4003  &  1   \\
      ZP$_{\mathrm{2hyb}}$  & 0.00454 &  0.0045 &  0.4061   &  1.0047  \\
      ZP$_{\mathrm{3grat}}$ & 0.00141 &  0.0015 &   0.4056   &  1.0032 \\
    \end{tabular}
    \caption{The normalized mean square error (nmse), normalized mean square error after morphological opening (nmse-o), relative efficiency, 1/c, and mean power $\|W\|^2$ for the URA, holographic, and bandwidth limited random pytchography zone plates.}
    \label{tab:my_label}
\end{table}
 
\newcommand{\fwidth}{0.24\textwidth}

\begin{figure*}[h]
    \begin{center}
    \includegraphics[width=\fwidth]{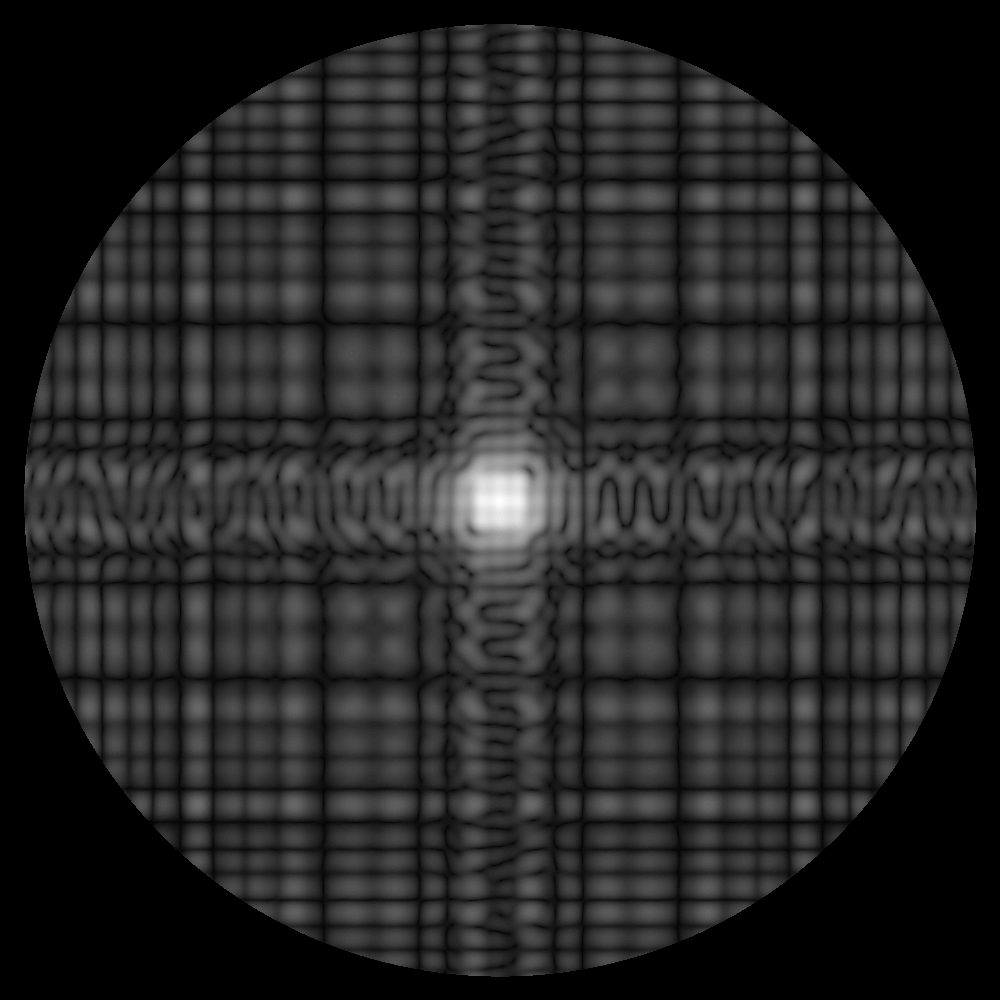}
    \includegraphics[width=\fwidth]{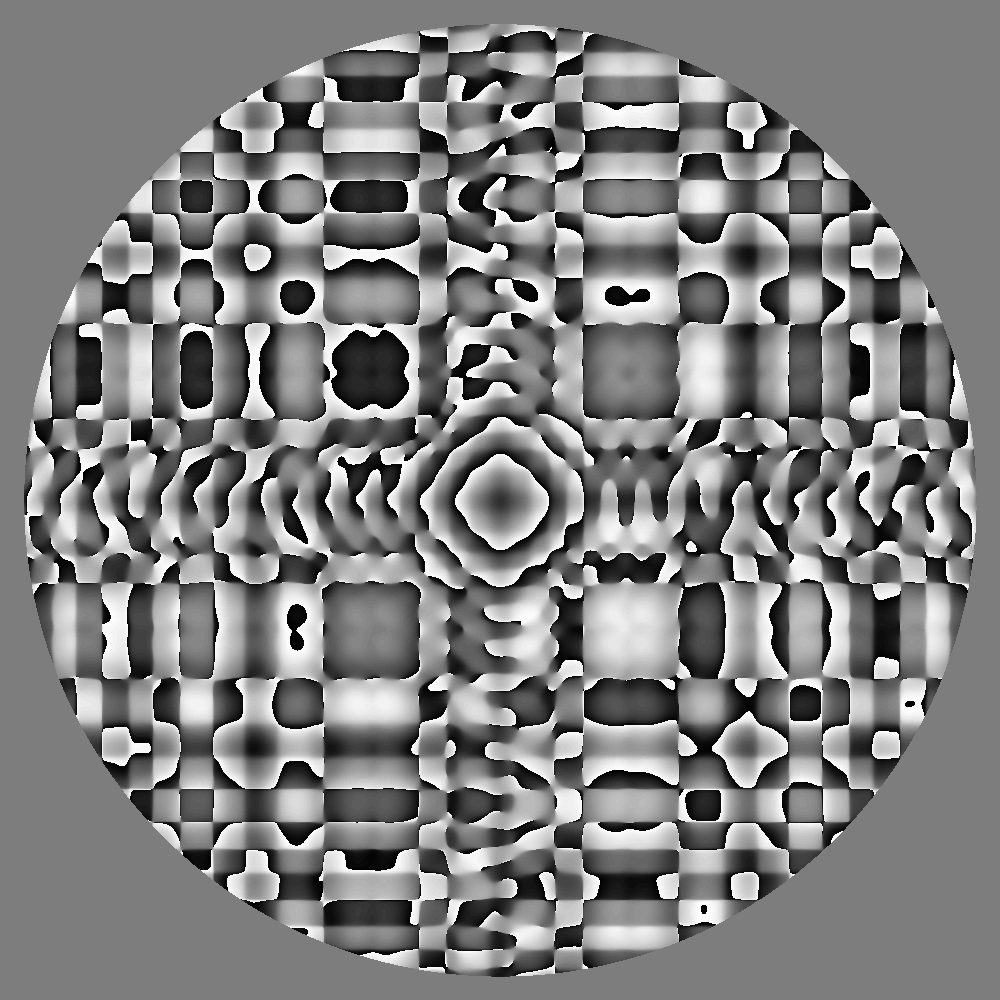}
    \includegraphics[width=\fwidth]{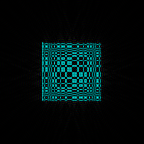}
    \includegraphics[width=\fwidth]{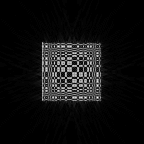}
    \\
    \includegraphics[width=\fwidth]{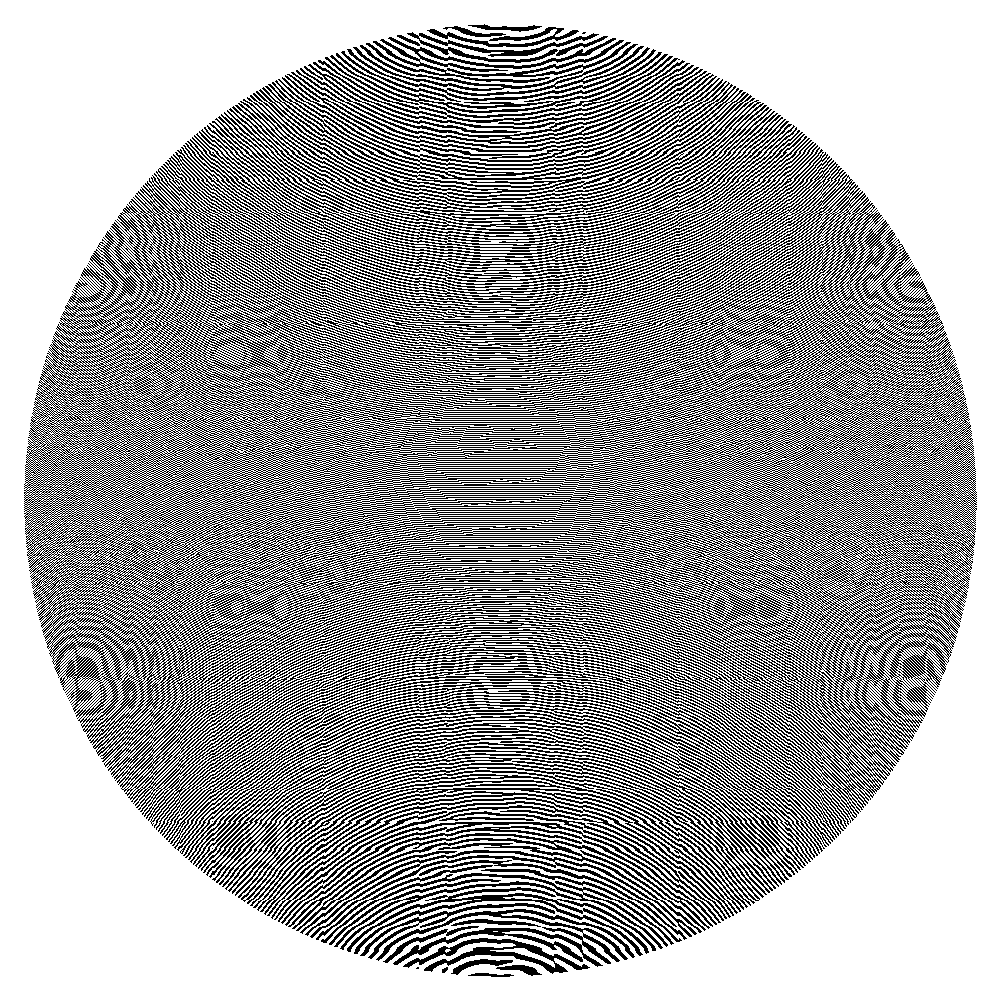}
    \includegraphics[width=\fwidth]{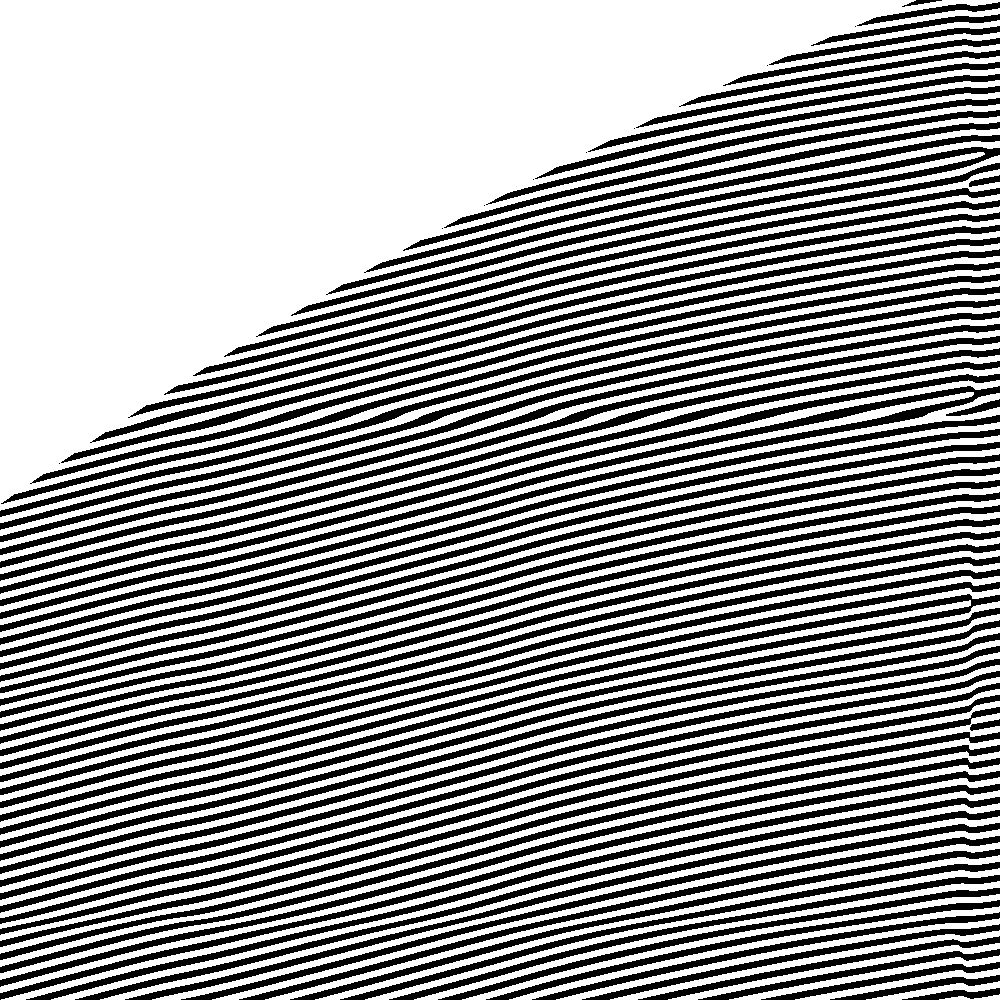}
    \includegraphics[width=\fwidth]{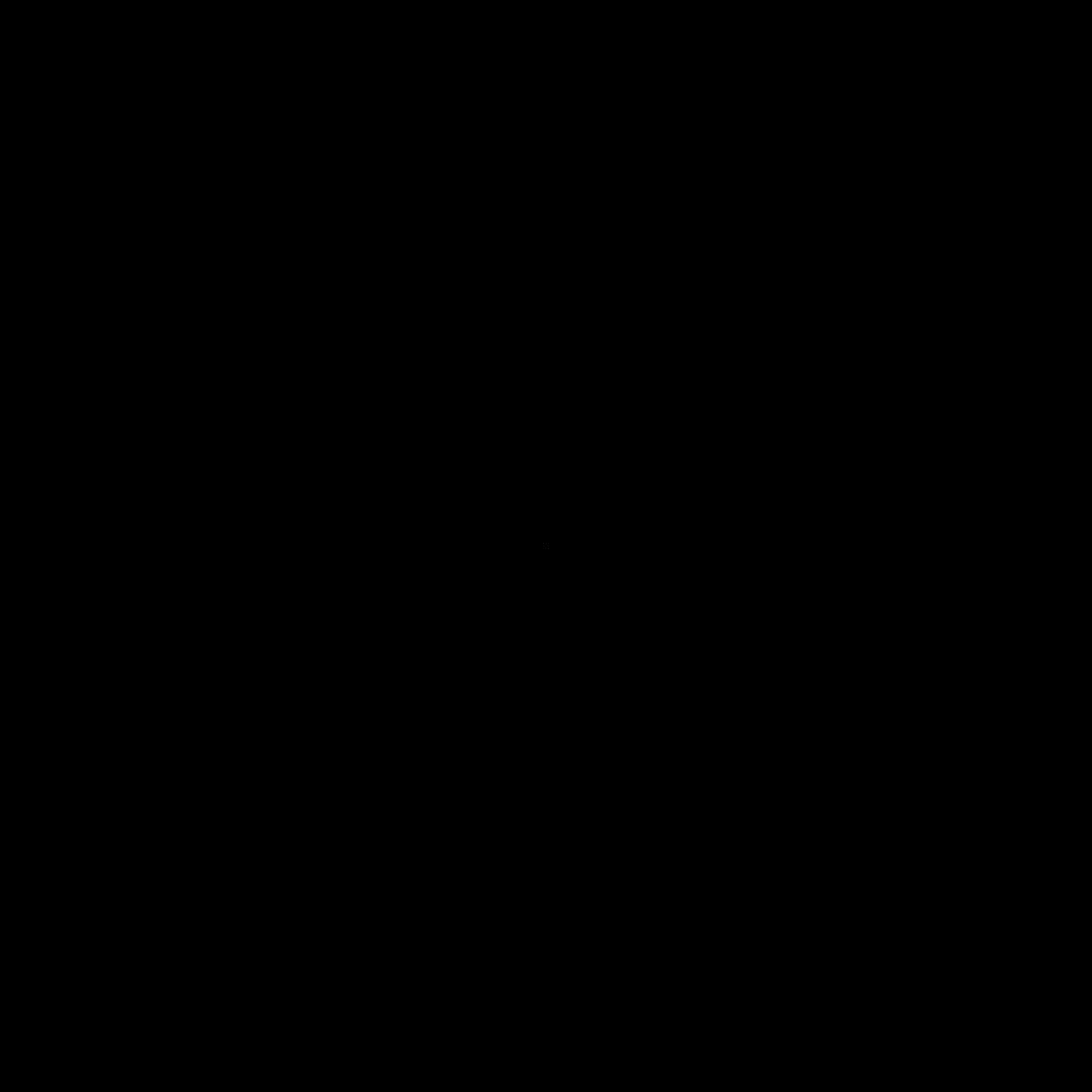}
    \includegraphics[width=\fwidth]{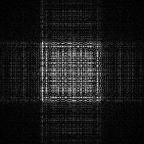}\\\
    \includegraphics[width=\fwidth]{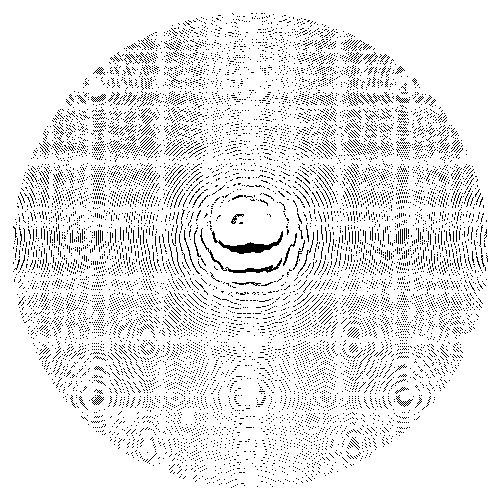}
    \includegraphics[width=\fwidth]{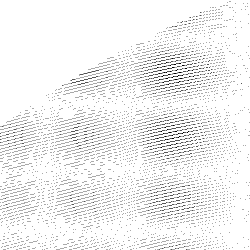}
    \includegraphics[width=\fwidth]{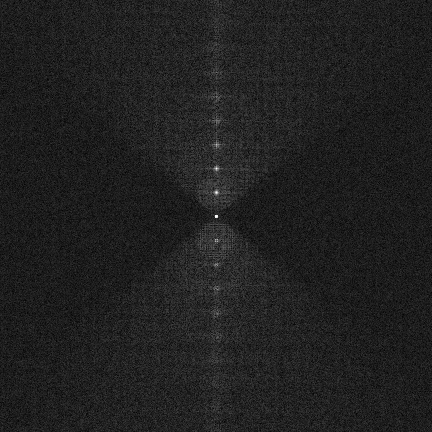}
    \includegraphics[width=\fwidth]{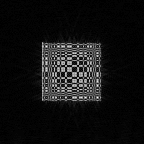}\\
    \includegraphics[width=\fwidth]{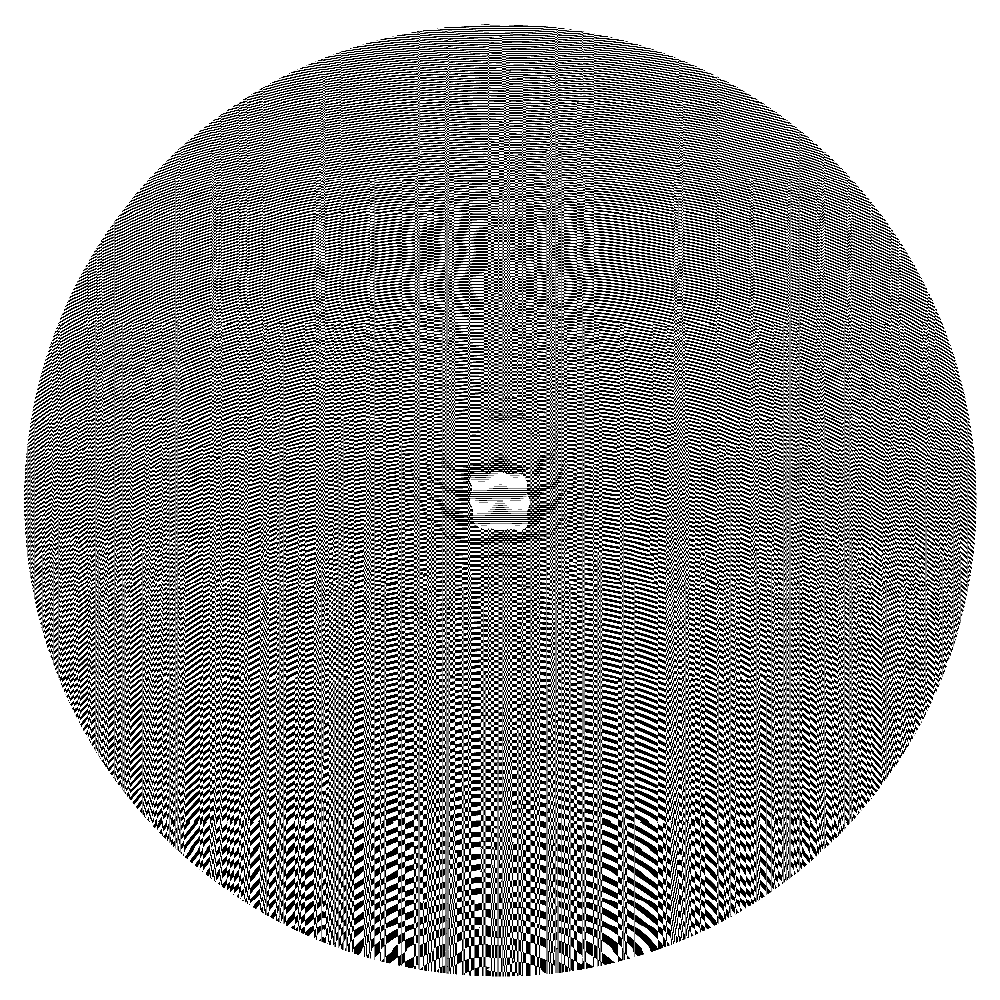}
    \includegraphics[width=\fwidth]{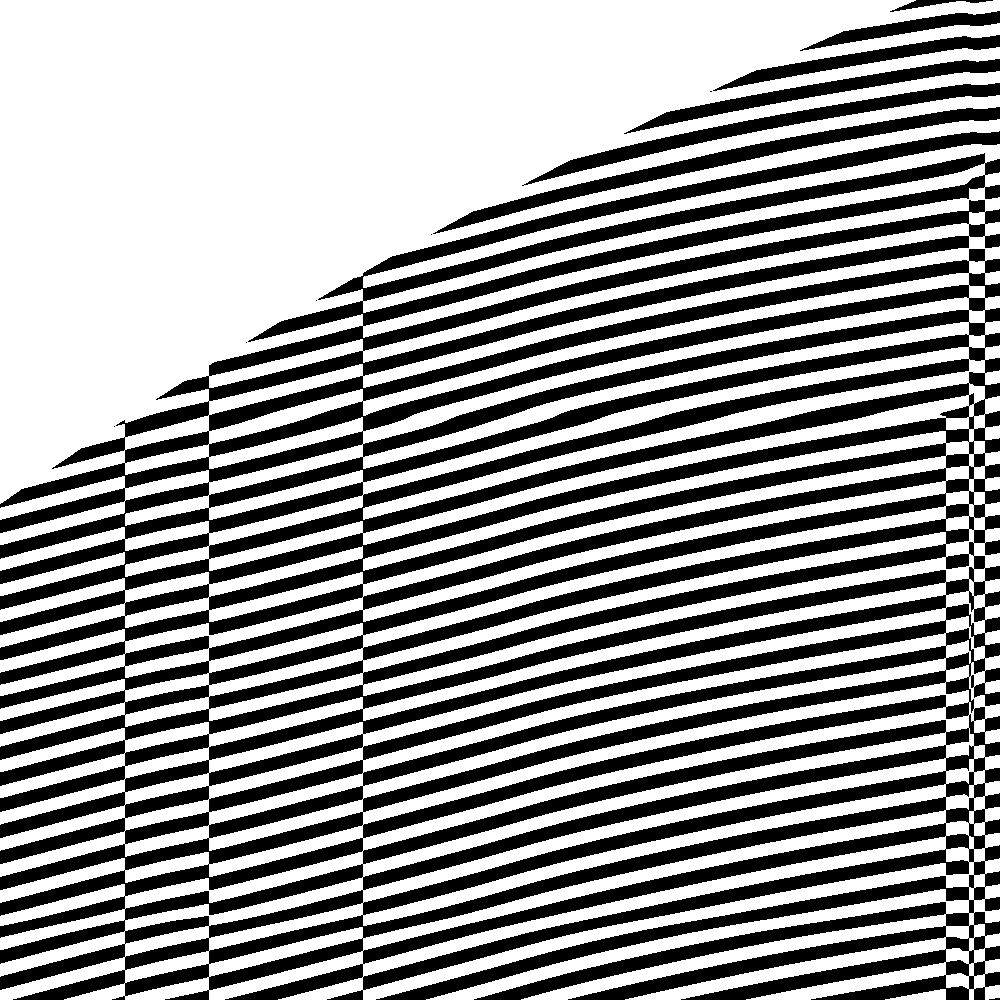}
    \includegraphics[width=\fwidth]{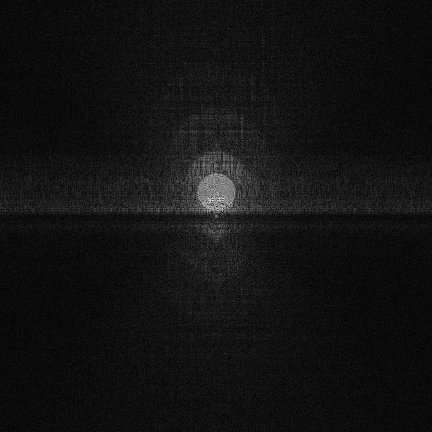}
    \includegraphics[width=\fwidth]{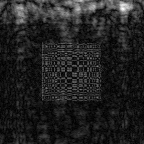} \\
     \includegraphics[width=\fwidth]{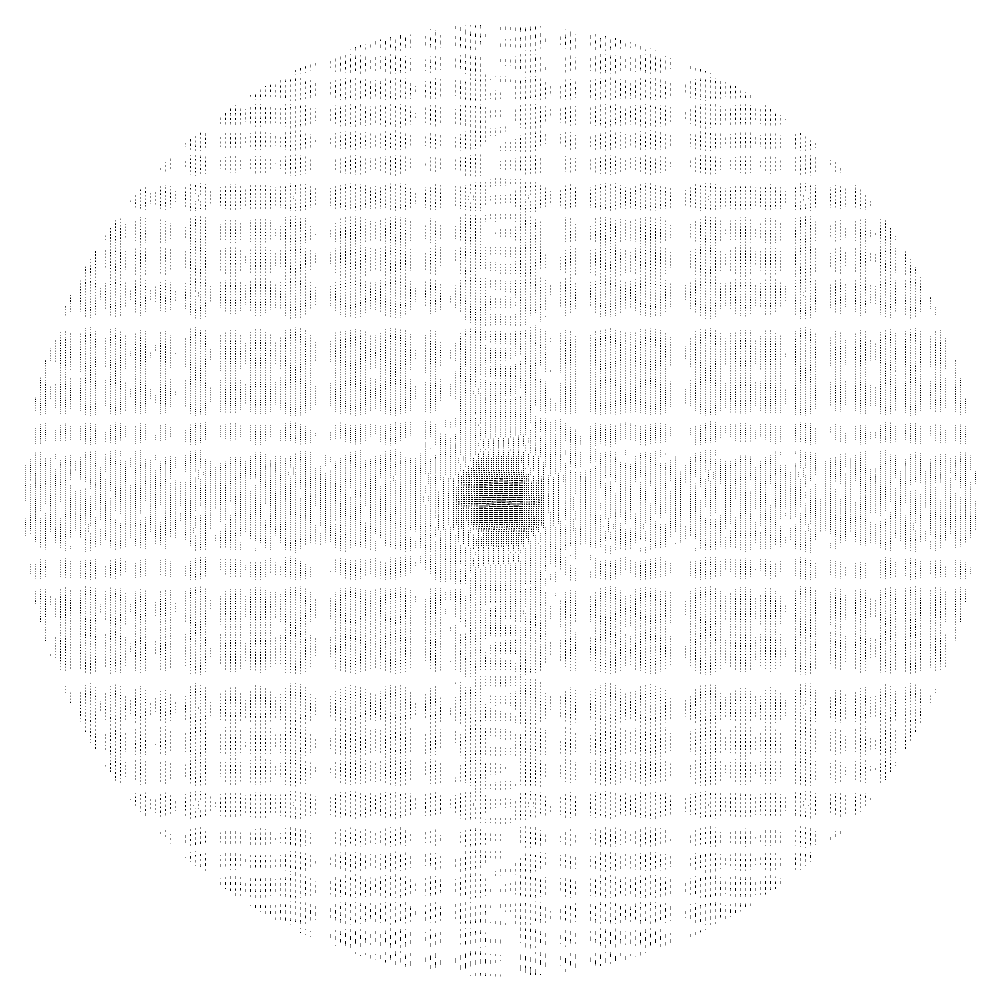}
     \includegraphics[width=\fwidth]{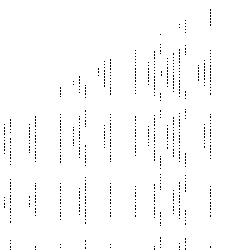}
     \includegraphics[width=\fwidth]{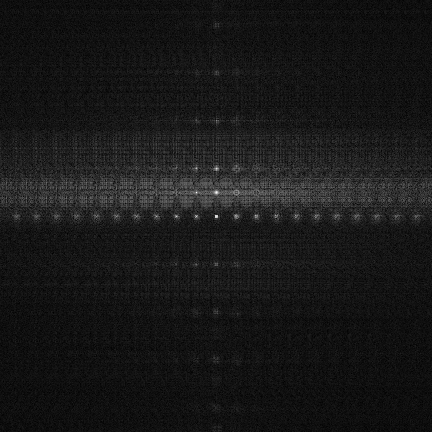}
     \includegraphics[width=\fwidth]{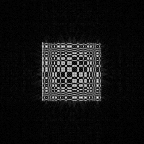}
    \end{center}
    \caption{Uniformly redundant array illumination.  Top row: desired amplitude, phase at the lens, and at the
sample plane complex-HSV representation and amplitude. Following Rows, from top to bottom, $W_{\mathrm{ZP}_0}$, $W_{\mathrm{ZP}_{1dc}}$, $W_{\mathrm{ZP}_{2hyb}}$, and $W_{\mathrm{ZP}_{3grat}}$.
    Columns, from left to right: the zone plate, magnified portion of zone plate, focal plane overview, and the focal spot.}
    \label{fig:URA}
\end{figure*}

\begin{figure*}[h]
    \begin{center}
    \includegraphics[width=\fwidth]{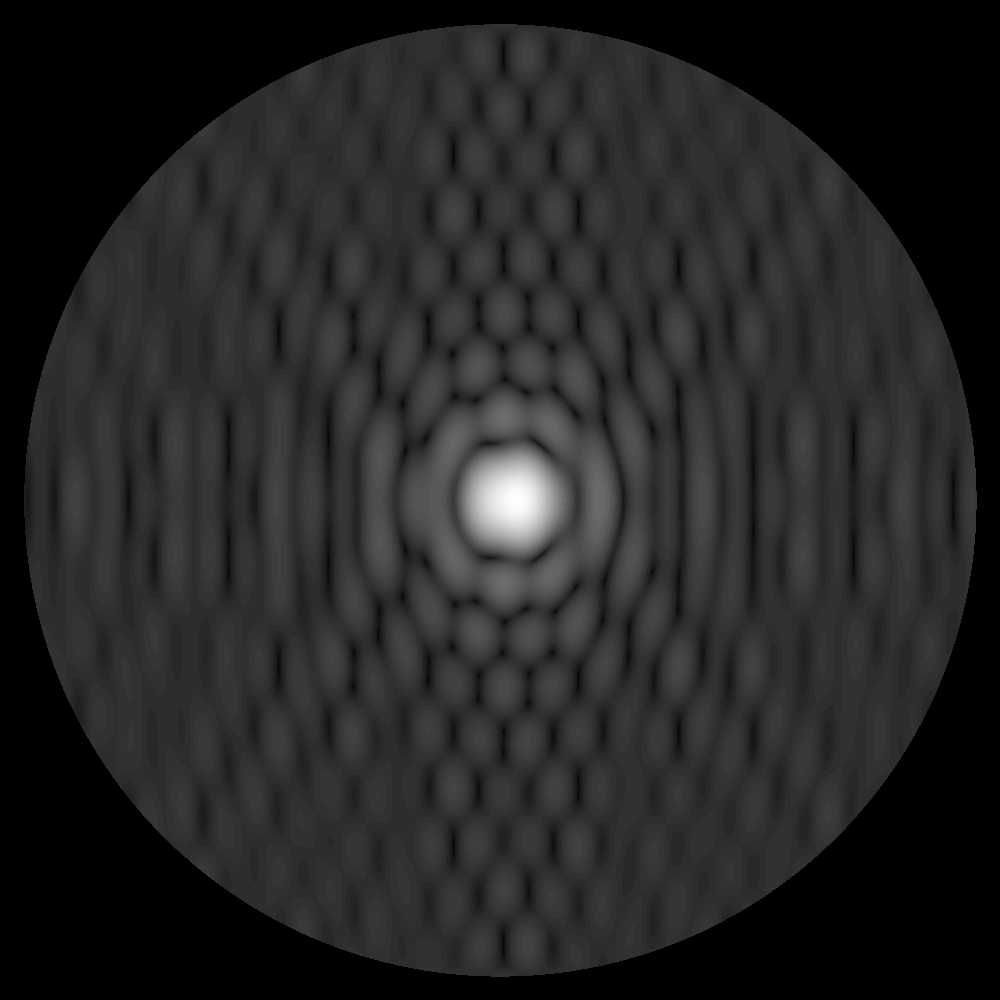}
    \includegraphics[width=\fwidth]{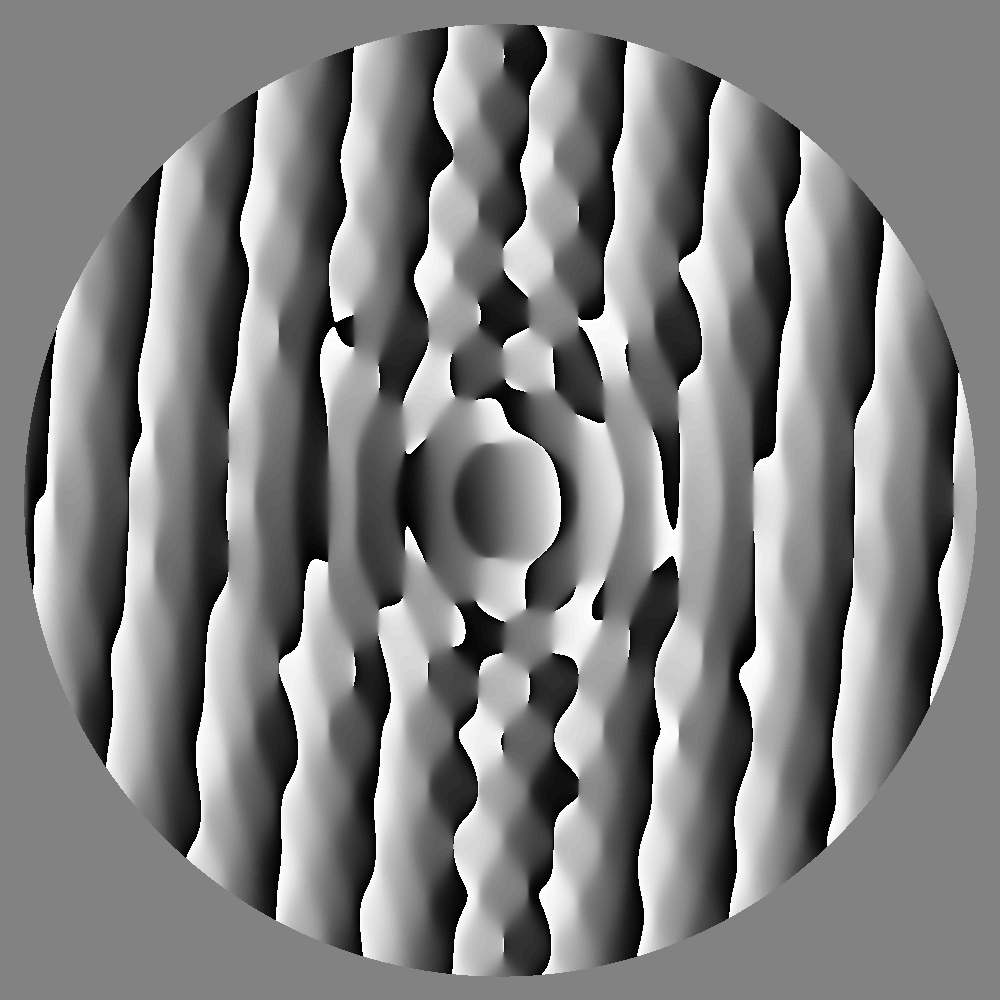}
    \includegraphics[width=\fwidth]{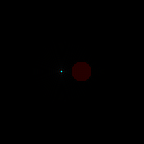}
    \includegraphics[width=\fwidth]{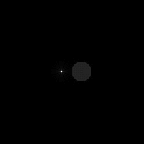}
    \\
    \includegraphics[width=\fwidth]{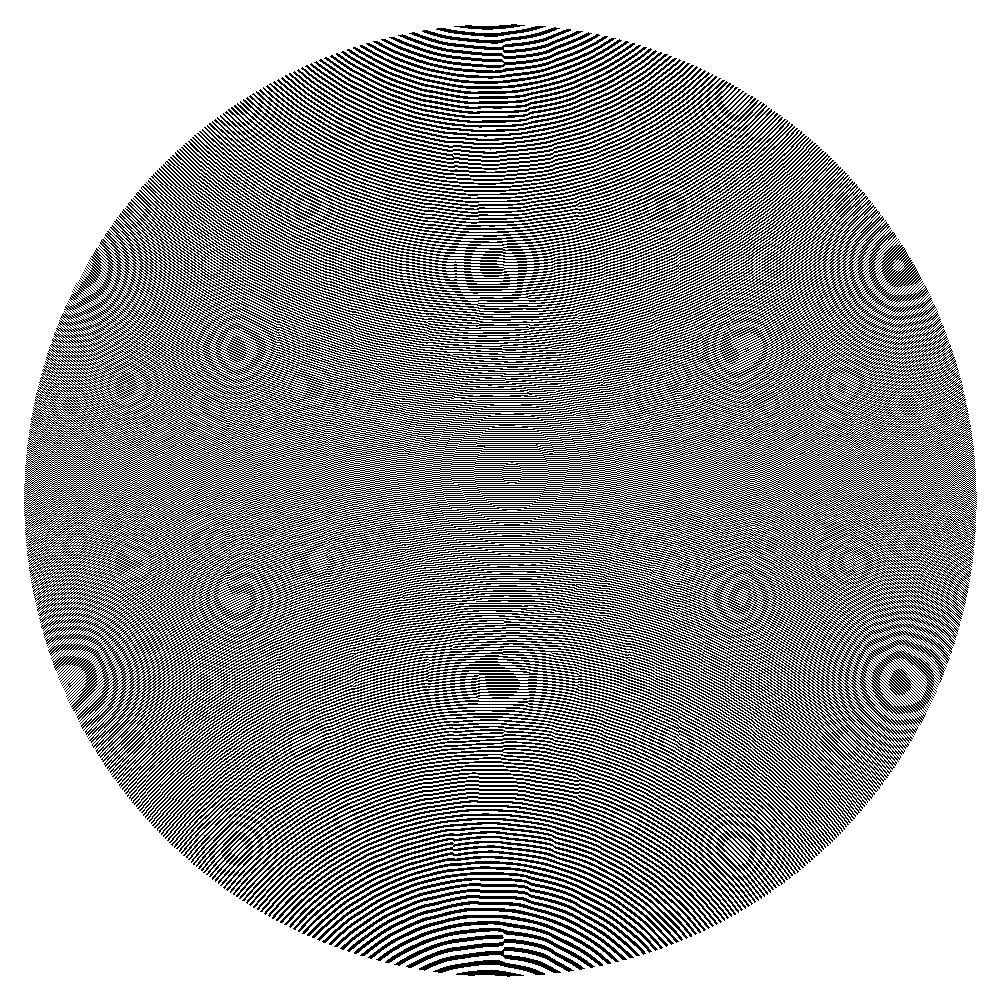}
    \includegraphics[width=\fwidth]{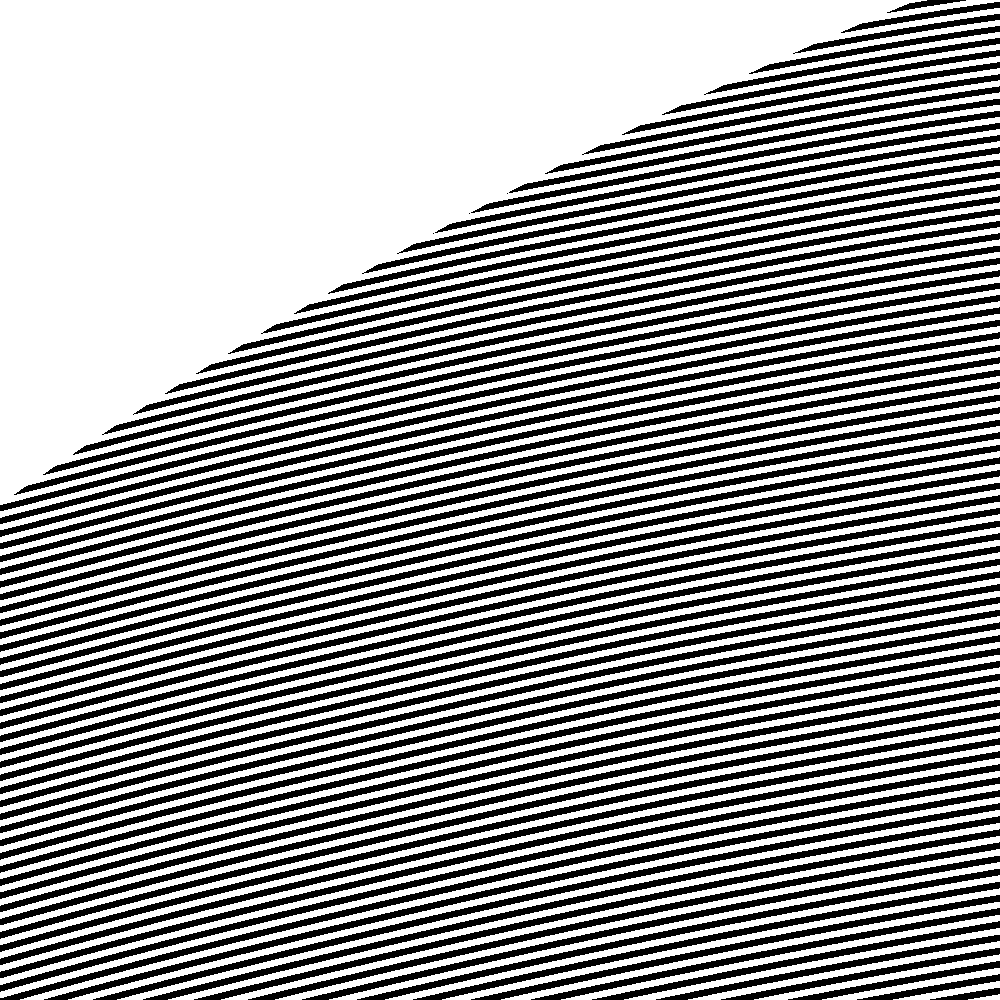}
    \includegraphics[width=\fwidth]{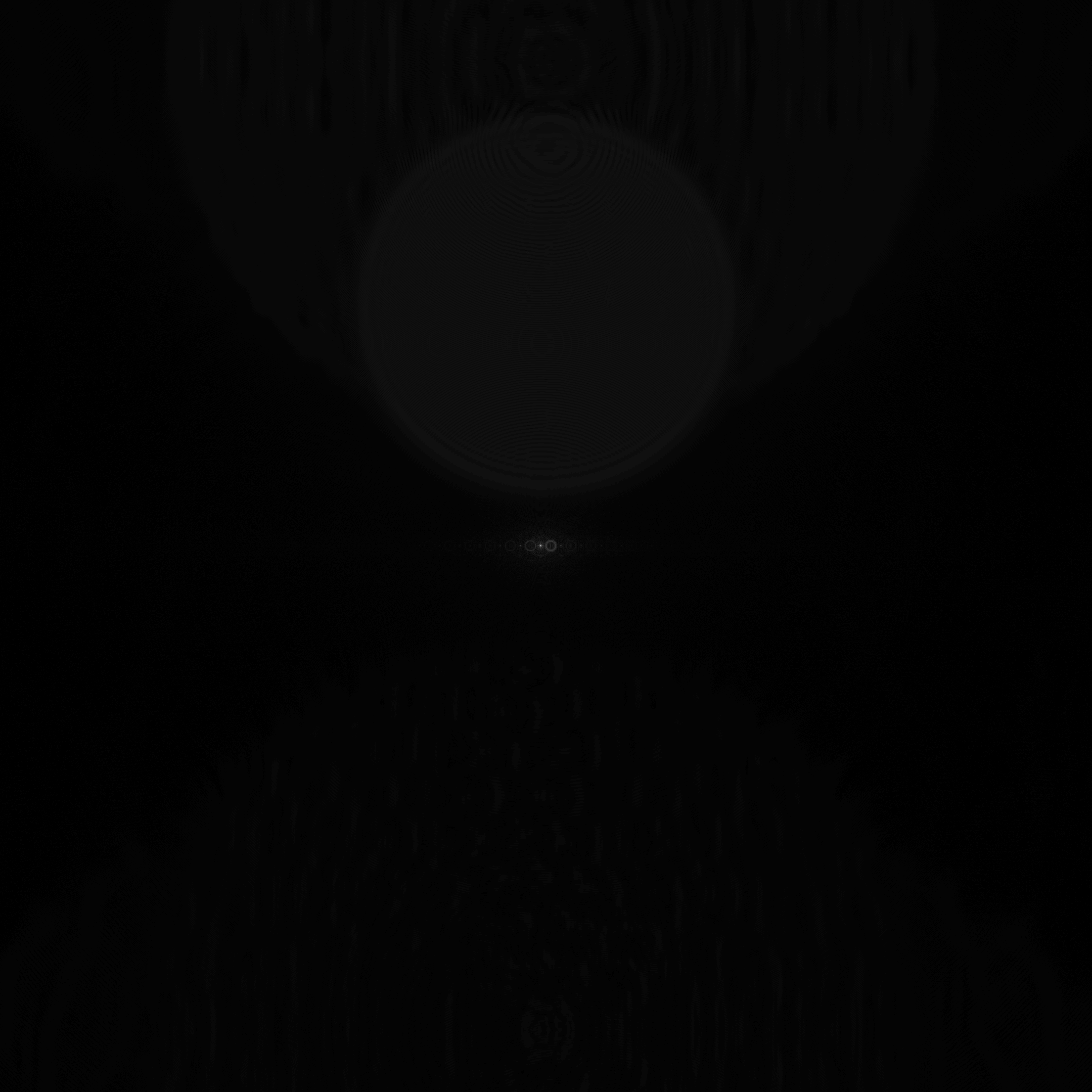}
    \includegraphics[width=\fwidth]{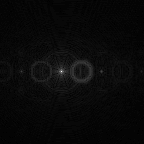}\\
    \includegraphics[width=\fwidth]{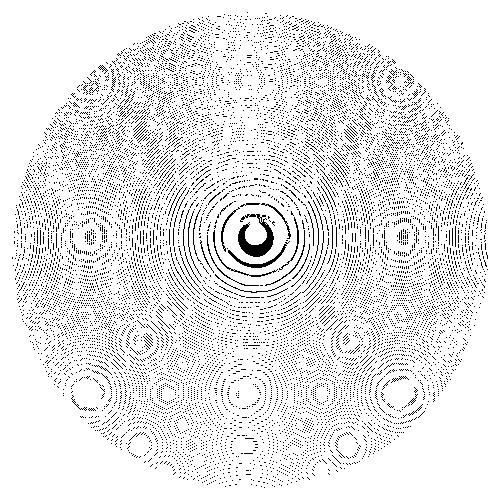}
    \includegraphics[width=\fwidth]{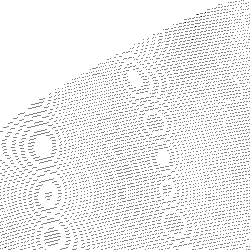}
    \includegraphics[width=\fwidth]{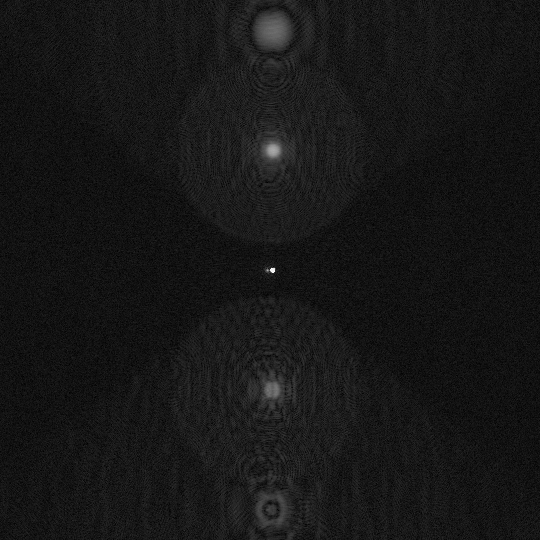}
    \includegraphics[width=\fwidth]{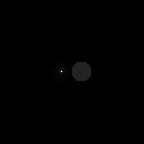}\\
    \includegraphics[width=\fwidth]{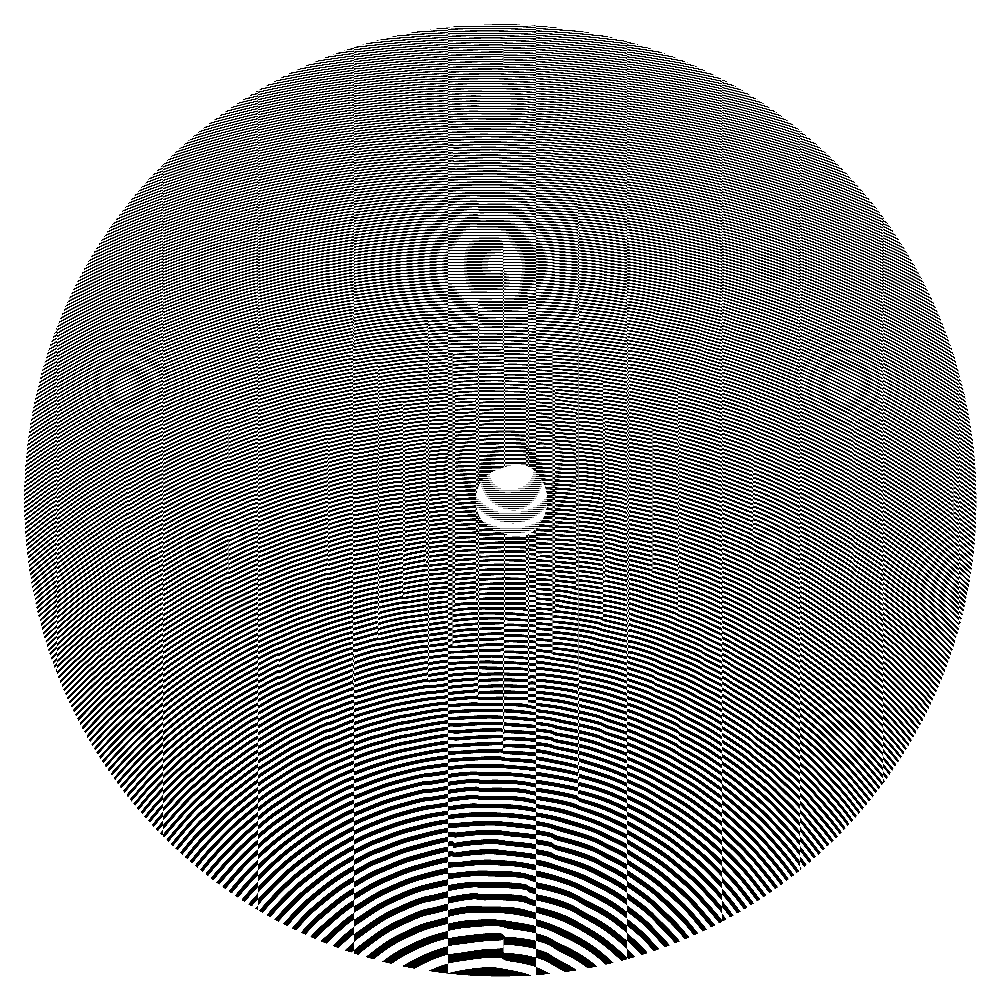}
    \includegraphics[width=\fwidth]{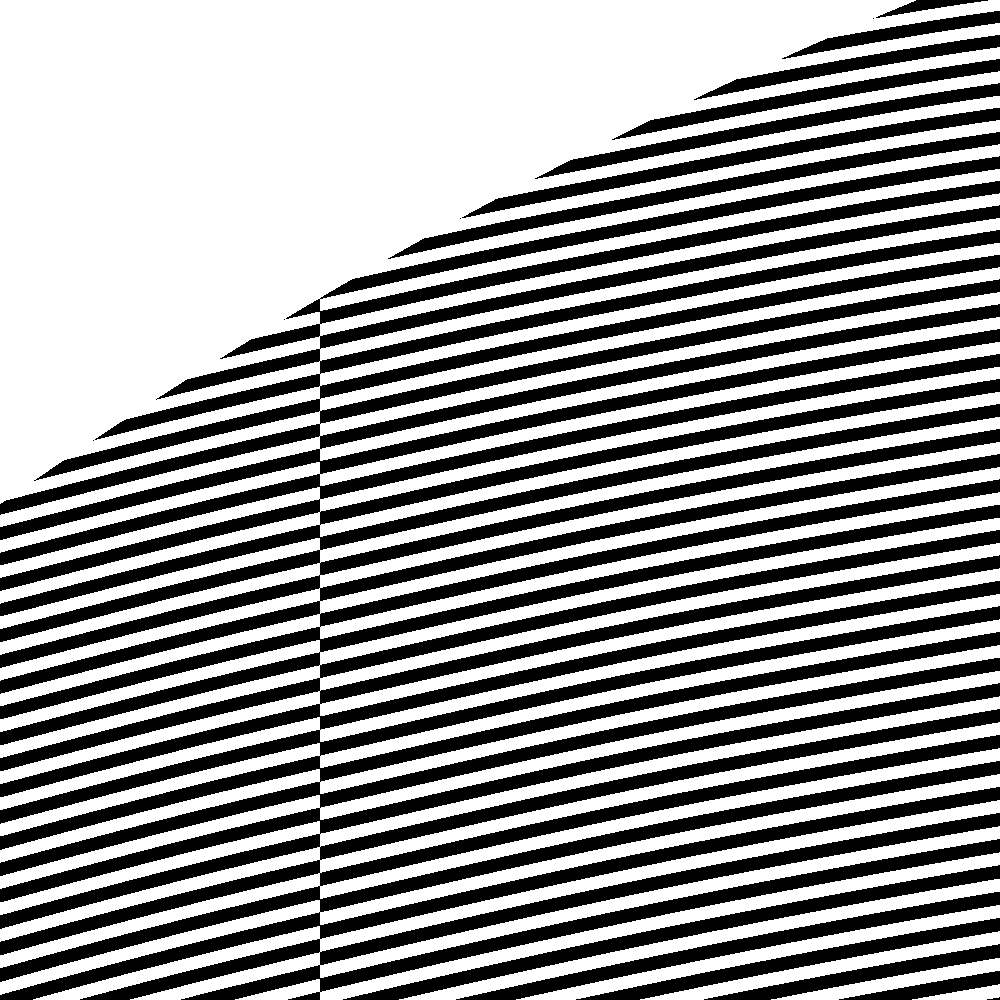}
    \includegraphics[width=\fwidth]{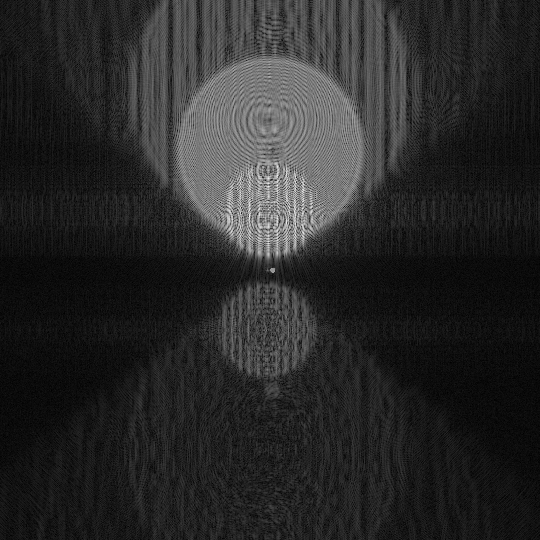}
    \includegraphics[width=\fwidth]{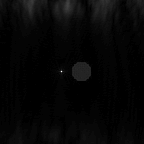}\\
    \includegraphics[width=\fwidth]{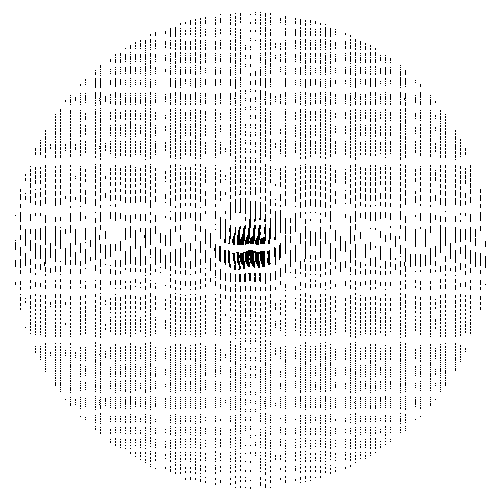}
    \includegraphics[width=\fwidth]{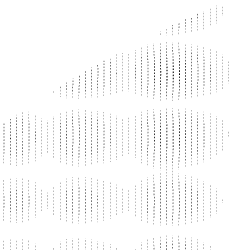}
    \includegraphics[width=\fwidth]{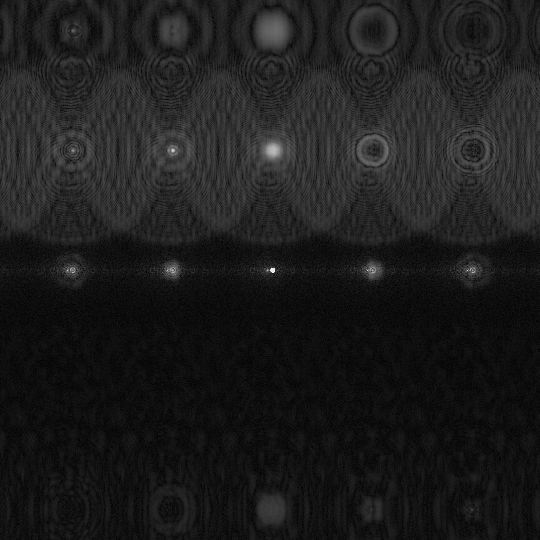}
    \includegraphics[width=\fwidth]{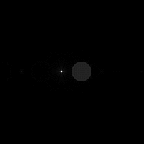}
    \end{center}
    \caption{Off-axis holographic illumination.  Top row: desired amplitude, phase at the lens, and at the
sample plane. Following Rows, from top to bottom, $W_{\mathrm{ZP}_0}$, $W_{\mathrm{ZP}_{1dc}}$, $W_{\mathrm{ZP}_{2hyb}}$, and $W_{\mathrm{ZP}_{3grat}}$.
    Columns, from left to right: the zone plate, magnified portion of zone plate, focal plane overview, and the focal spot.}
    \label{fig:holo}
\end{figure*}
 
  \begin{figure*}[h!]
    \begin{center}
    \includegraphics[width=\fwidth]{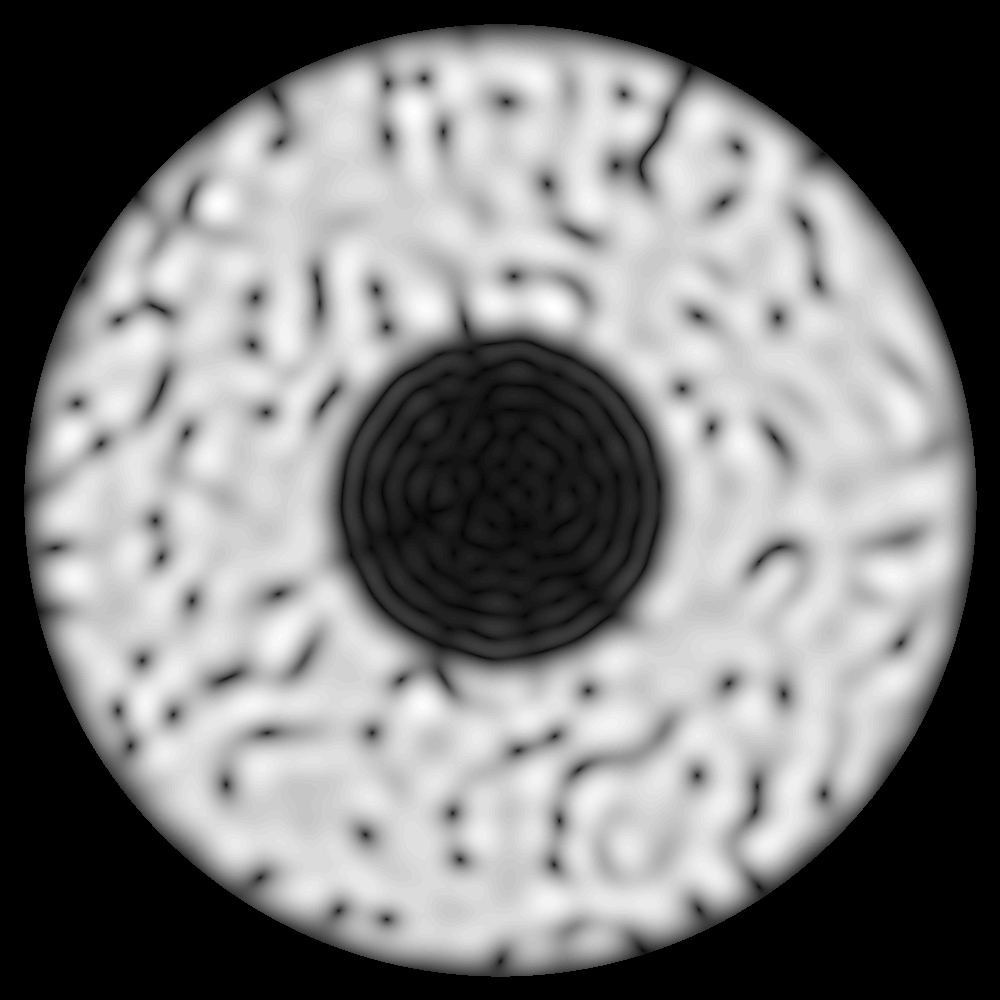}
    \includegraphics[width=\fwidth]{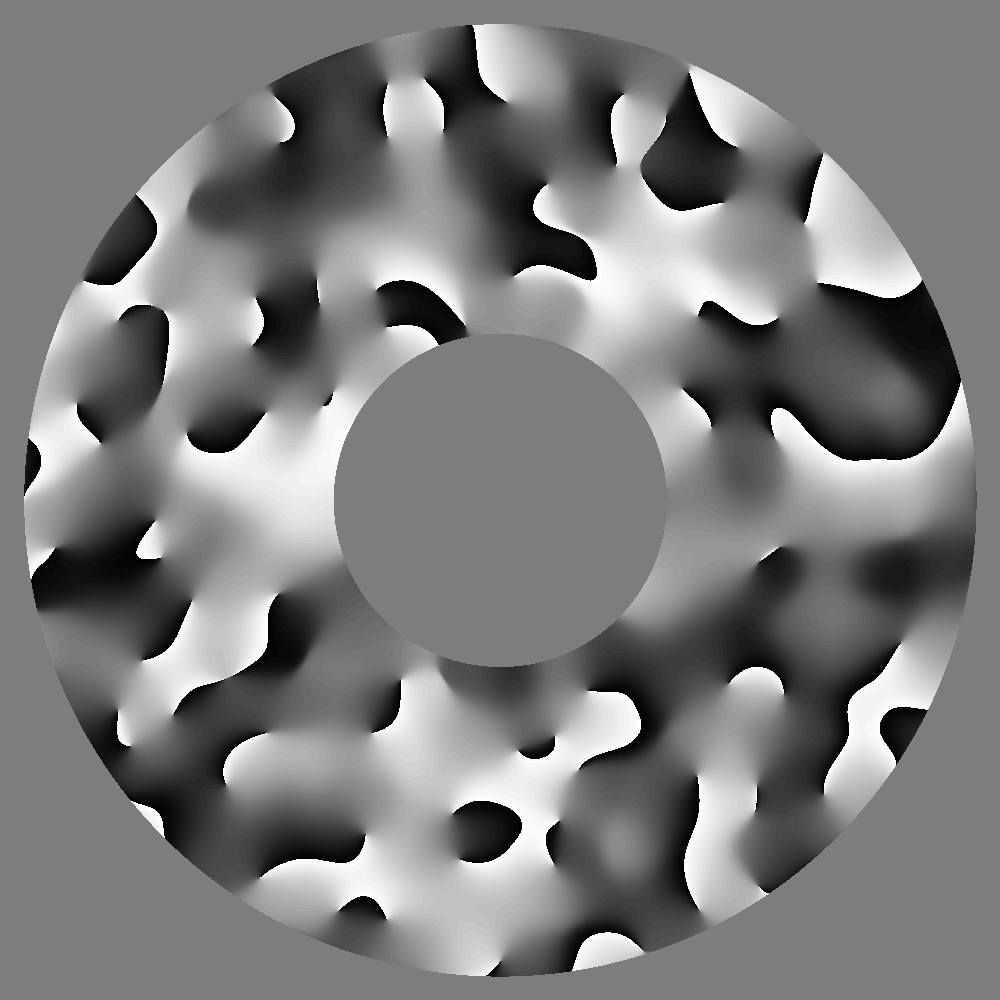}
    \includegraphics[width=\fwidth]{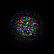} 
    \includegraphics[width=\fwidth]{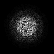}
    \\
    \includegraphics[width=\fwidth]{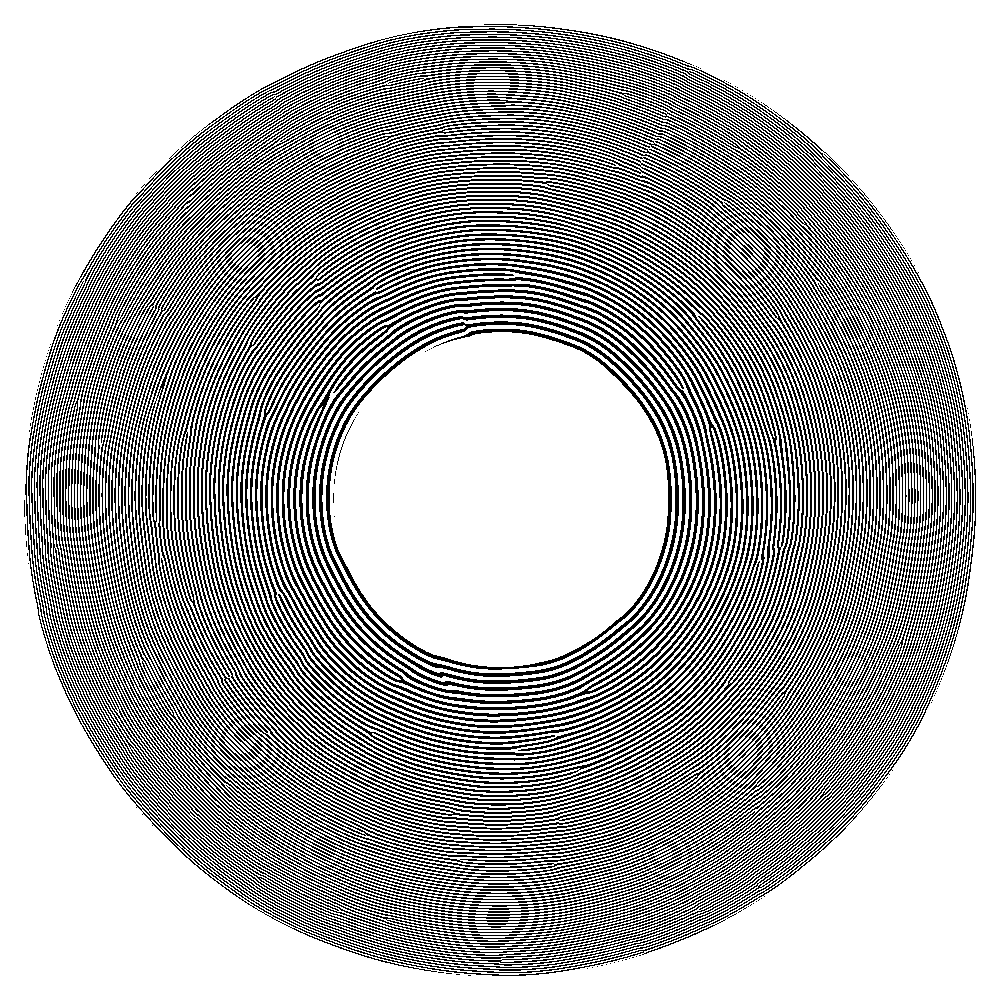}
    \includegraphics[width=\fwidth]{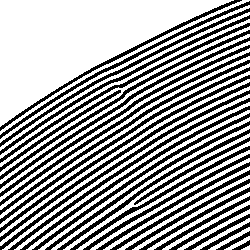}
    \includegraphics[width=\fwidth]{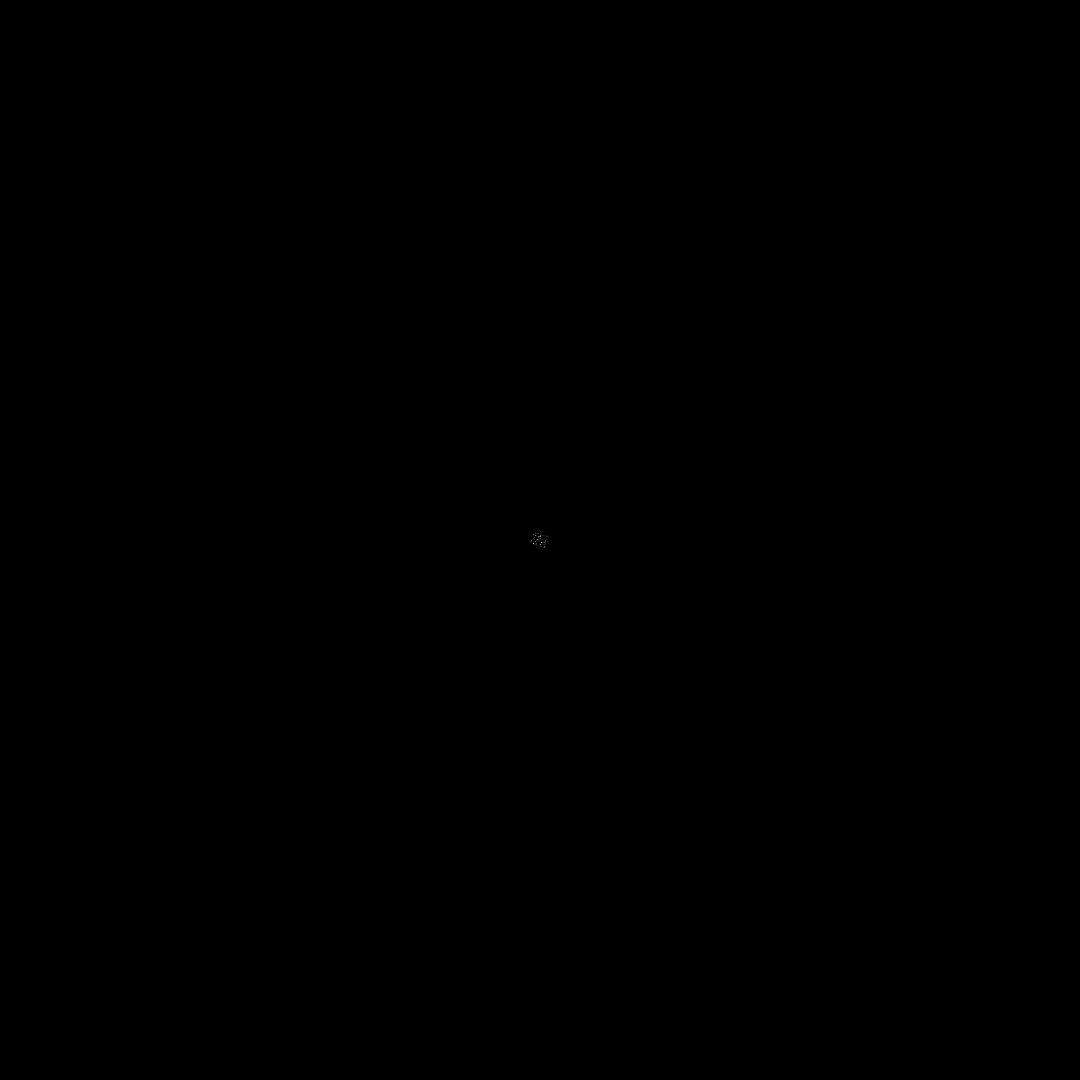}
    \includegraphics[width=\fwidth]{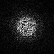}\\
    \includegraphics[width=\fwidth]{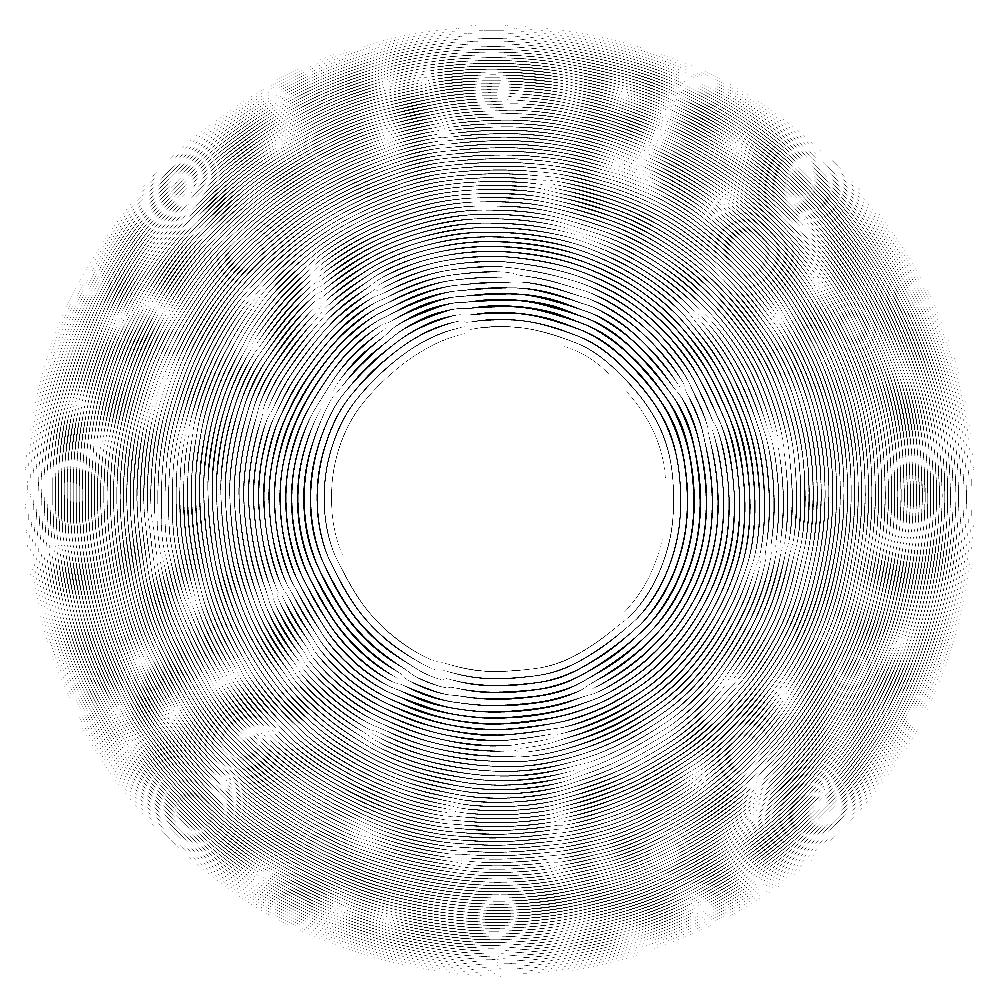}
    \includegraphics[width=\fwidth]{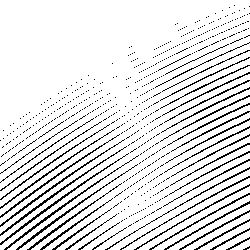}
    \includegraphics[width=\fwidth]{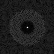}
    \includegraphics[width=\fwidth]{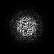}\\
    \includegraphics[width=\fwidth]{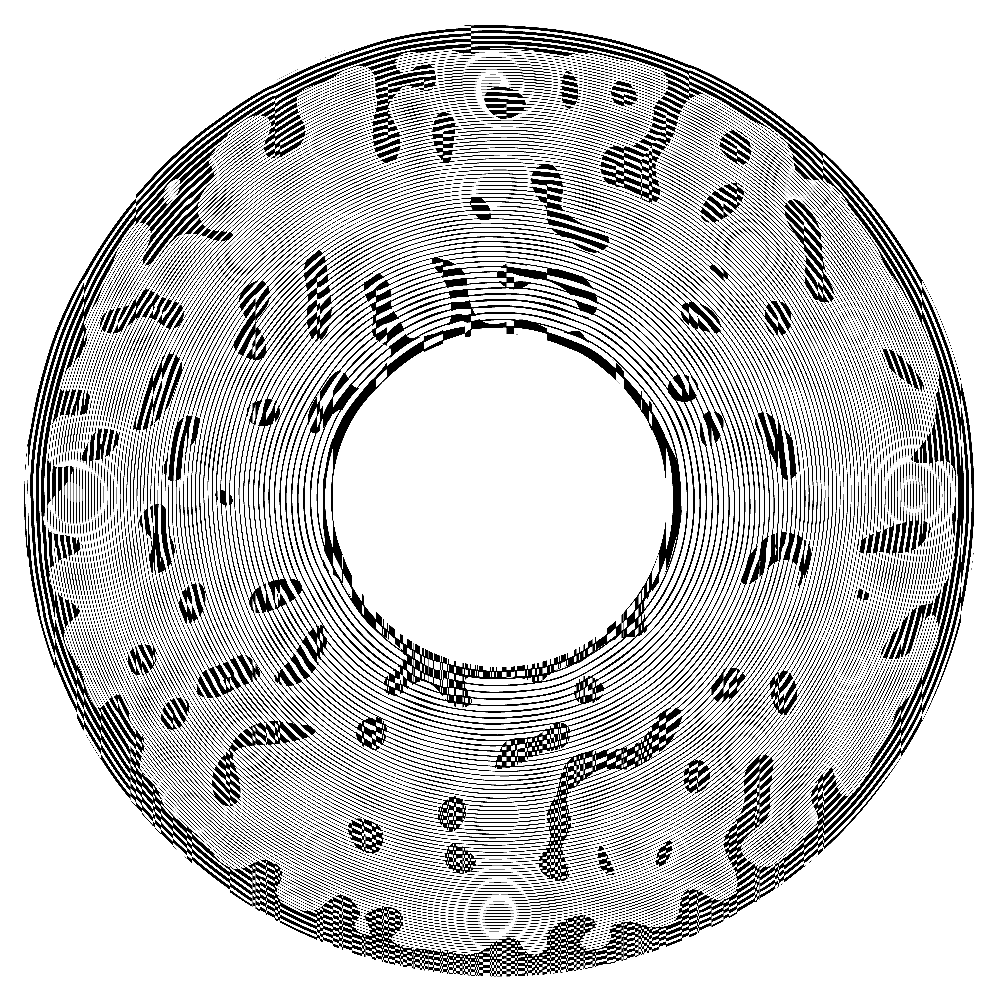}
    \includegraphics[width=\fwidth]{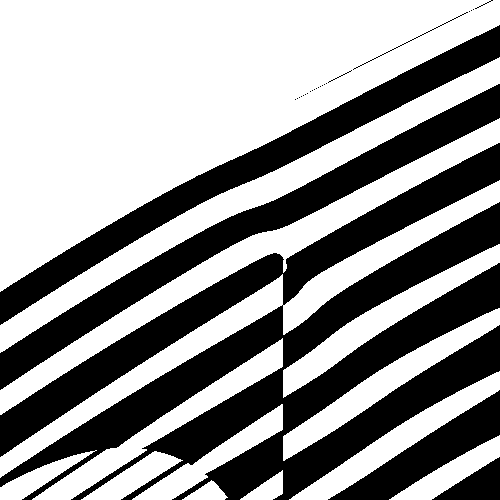}
    \includegraphics[width=\fwidth]{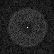}
    \includegraphics[width=\fwidth]{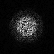}\\
    \includegraphics[width=\fwidth]{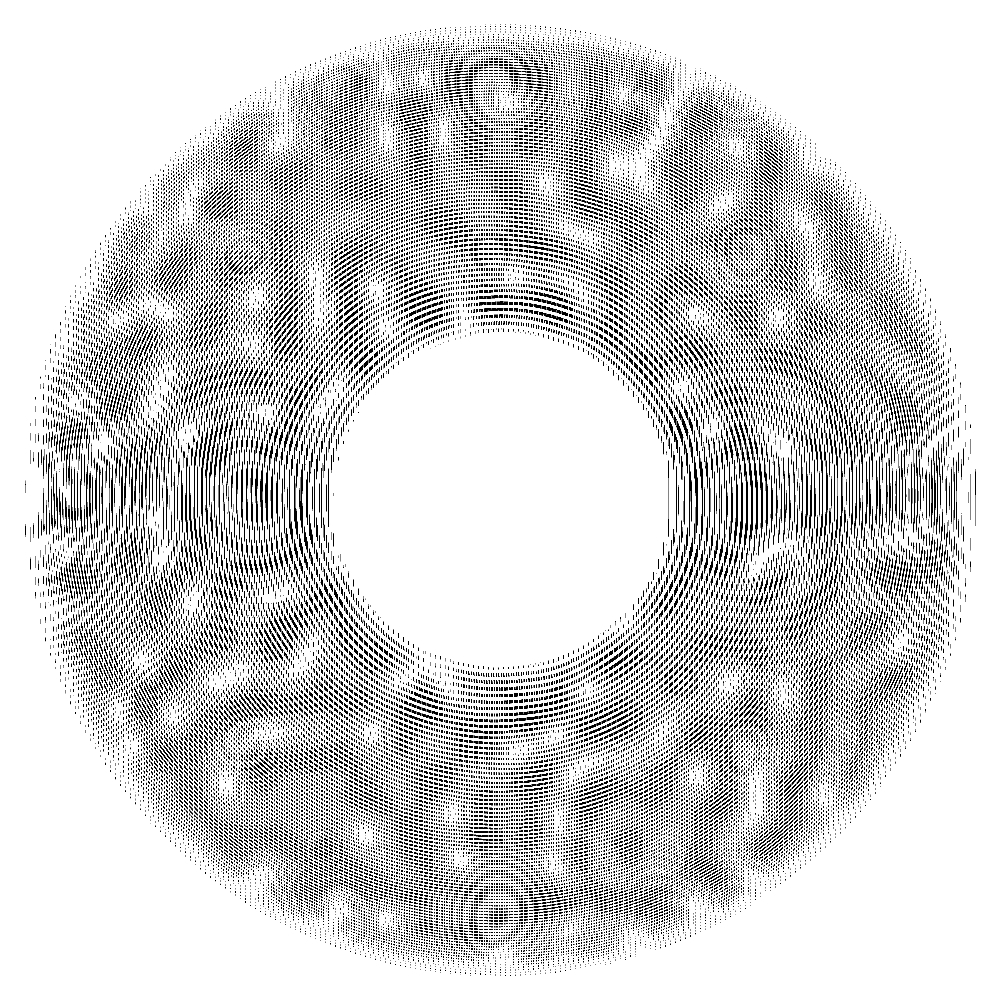}
    \includegraphics[width=\fwidth]{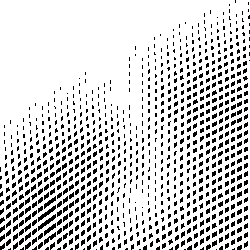}
    \includegraphics[width=\fwidth]{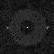}
    \includegraphics[width=\fwidth]{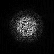}
    \end{center}
    \caption{Band Limited Random illumination. Top row: desired amplitude, phase at the lens, and at the sample represented in HSV color format with Hue=phase, Value=amplitude, and Saturation=1, and amplitude on the right. 
    Following Rows, from top to bottom, $W_{\mathrm{ZP}_0}$, $W_{\mathrm{ZP}_{1dc}}$, $W_{\mathrm{ZP}_{2hyb}}$, and $W_{\mathrm{ZP}_{3grat}}$.
    Columns, from left to right: the zone plate, magnified portion of zone plate, focal plane overview, and the focal spot.}
    \label{fig:BLR}
\end{figure*}

Simulations for three practical examples using two off-axis and one on-axis zone plate design were performed: structured illumination with a uniformly redundant array \cite{fenimore1978coded} used in \cite{marchesini2008massively}, an off-axis holographic illuminator with a bright high resolution reference beam and a large illuminated sample beam, and a bandwidth limited randomly structured illumination beam for ptychography\cite{marchesini2016alternating}.  

In each of these cases, the classical technique of thresholding the phase component was compared to each of the three zone plate design techniques described earlier. Figure \ref{fig:URA} shows simulations for the case of the uniformly redundant array illuminator.  Figure \ref{fig:holo} shows simulations for the case of the off-axis holographic illuminator.  Figure \ref{fig:BLR} shows simulations for the case of the randomly structured illumination for ptychography.  In each of the figures, the first row shows the desired phase and amplitude in the lens and focal planes, the complex illumination is represented in HSV (Hue=phase, Saturation=1, Value=amplitude).  The second row shows results from the classical technique, the third row shows results from $W_{\mathrm{ZP}_{1dc}}$ with duty cycle modulation, the fourth row shows results from $W_{\mathrm{ZP}_{2hyb}}$ with duty cycle modulation combined with higher harmonic order modulation, and the fifth row shows results from the $W_{\mathrm{ZP}_{3grat}}$ with amplitude modulation by a grating.   For each technique, from left to right, the zone plate design, a magnified portion of the zone plate, the focal plane overview, and the focal spot selected after use of an order sorting aperture are depicted.  

The fidelity can be described by the quantization error, which can be determined using the normalized mean square error (nmse):
\begin{equation} \label{eq:nmse}
\mathrm{nmse}= \min_{c} \frac{\| w_\mathrm{truth} - c w_\mathrm{ZP} \|^2_{\mathrm{OSA}} }{\| w_\mathrm{truth}  \|^2_{\mathrm{OSA}}},
\end{equation}
 where $\mathrm{OSA}$ is a small region around the optical axis, $c$ is a scalar factor that minimizes the difference with the desired illumination $w_\mathrm{truth}$. The smaller the nmse, the higher the fidelity of the focal spot, and the larger the nmse, the lower the fidelity of the focal spot. 

While nmse considers the fidelity to the original design itself, it does not take into account possible fabrication limitations or errors.  In order to evaluate the robustness of each of the design to fabrication error we apply a morphological 
opening \cite{vincent1993morphological}  - the dilation of the erosion of an image - to the ideal binary zone plate design with a structuring element disk of 1 pixel radius. The result of this is referred to as nmse-o.  The smaller the nmse-o, the higher the fidelity of the focal spot in presence of simulated fabrication error, and the higher the nmse-o, the lower the fidelity of the focal spot in presence of simulated fabrication error. While the evaluation of the possible fabrication errors depend on the specific lithographic process, x-ray energy, and experimental geometry, the $W_{\mathrm{ZP}_{2hyb}}$ design maintains the fidelity better compared with the other designs after a morphological opening operation.

 The scaling factor between designed illumination $w_\mathrm{truth}$ normalized so that $\max | {\cal F} w_\mathrm{truth} |= 1$ and the zone plate $w_\mathrm{ZP}$ is given by $1/c$. A measure of relative diffraction efficiency is described by the scaling factor between the first order diffraction efficiency of $\mathrm{ZP}_0$ and each of the other zone plates. Because these new zone plates modulate the amplitude contribution of the zone plate, the first order diffraction efficiency can only be less than that of $\mathrm{ZP}_0$, a binary zone plate with a 1:1 duty cycle, and it is related to the average power $\|W\|^2$ of the desired wavefront.  The absolute first order diffraction efficiency, however, will scale with not only the relative efficiency, but also with the specific zone plate material, zone plate thickness, and operational energy being used. Quantitative nmse, nmse-o, relative efficiency, and 1/c values for the cases of the uniformly redundant array, off-axis holography, and bandwidth limited random illuminations are found in Table \ref{tab:my_label}. 
 
 Without the consideration of fabrication limitations or errors, the highest fidelity focal spots are found with techniques ZP$_{\mathrm{1dc}}$ and ZP$_{\mathrm{3grat}}$.  However, fabrication challenges associated with producing such small linewidths could prevent practical implementation, and while the evaluation of the possible fabrication errors depends on the specific lithographic process, the  ZP$_{\mathrm{2hyb}}$ design maintains fidelity better in the presence of lithographic limitations compared with the other designs after a morphological opening operation. The relative efficiency of the lens becomes important for practical implementation of these techniques and varies proportionally to the power spectrum of the desired focal spot. 
 
 Depending on the application, the absolute efficiency of these designs may or may not be practically acceptable.  However, slight modifications to the design of the desired focal spot can result in better optimization and improvement of the relative efficiency, which is especially true for the first two examples of the URA and off-axis holography.  Alternately, depending on the specific experimental requirements, use of a source with several orders of magnitude more flux could compensate for lower zone plate efficiencies. For the case of the bandwidth limited random ptychography zone plate, where efficiency has been optimized, we can estimate absolute efficiency for a typical experiment.  For the case of a synchrotron-based experiment, the diffraction efficiency of a zone plate can be on the order of 2 - 40 percent, depending on experimental needs and fabrication capabilities\cite{VilaComamala2011zonedoubledzp, chang2014macezp, Mohacsi2015doublezp}.  An estimate of absolute efficiency for the case of the bandwidth limited random ptychography zone plate would then be given by the product of the relative efficiency and the diffraction efficiency, approximately 0.8 - 16 percent, which is well within experimentally acceptable limits\cite{morrison2018x,chen2018coded}.  In the case of an XFEL experiment, assuming the use of a diamond-based zone plate where efficiencies have been shown to be on order of 2-13 percent \cite{david2011diamondzp}, for pure diamond - diamond/Ir zone plates, respectively, implementation of a similar zone plate for XFELs would yield absolute efficiencies of approximately 0.8-5.2 percent, also within experimentally acceptable limits.

\section{Conclusions}
The ability to arbitrarily shape coherent x-ray wavefronts at new synchrotron and x-ray free electron facilities with these new optics will lead to advances in measurement capabilities and techniques that have been difficult to implement in the x-ray regime.  We proposed three types of binary zone plates designs optimized for wavefront shaping. These new lenses consider both phase and amplitude requirements and modify the standard zone plate using variations in the duty cycle, harmonic order, and/or orthogonal gratings.  These type of zone plates can generate nearly arbitrary illuminations with high fidelity.  However, due to the linewidth variations that result in very small linewidths, some of these zone plates may be difficult to fabricate.  This can be overcome by implementing a hybrid zone plate that utilizes patches of higher harmonic orders to account for the required amplitude variations while enabling fabrication within current lithographic limitations.  Three practical examples that can have on improvements in x-ray techniques were simulated: structured uniformly redundant array illumination, off-axis holographic imaging, and improved ptychographic imaging.  We envision further applications to enhance scientific studies such as flat top beams for pump-probe studies multiplexed beams for improved sensitivity in materials spectroscopy, among others.  As wavefront shaping capabilities with visible optics and spatial light modulating technology have led to numerous advances in the visible light regime, so is the potential for wavefront shaping in the x-ray regime.

\section{Acknowledgments}
 This work was partially funded by the Advanced Light Source and the Center for Applied Mathematics for Energy Research Applications, a joint ASCR-BES funded project within the Office of Science, US Department of Energy, under contract number DOE-DE-AC03-76SF00098.  This work was partially funded by the DOE Early Career Program and the Stanford Synchrotron Radiation Lightsource, SLAC National Accelerator Laboratory, which is supported by the U.S. Department of Energy, Office of Science, under Contract No. DE-AC02-76SF00515.  



\bibliography{citations}

\end{document}